\documentclass[12pt,twoside]{article}
\usepackage{fleqn,espcrc1}
\usepackage{graphicx}
\usepackage{floatflt}

\def\Journal#1#2#3#4{{#1} {\bf #2} (#4) #3}

\def\NPA{{\em Nucl. Phys.} A}
\def\PLB{{\em Phys. Lett.} B}

\newcommand{\pt}{$p_{\mathrm{T}}$}

\newcommand{\ccbar}{$c\bar{c}$}

\newcommand{\jpsi}{J/$\psi$}
\newcommand{\psip}{$\psi^\prime$}
\newcommand{\clearemptydoublepage}{\newpage{\pagestyle{empty}\cleardoublepage}}
\begin{document}

\begingroup
\thispagestyle{empty}
\baselineskip=14pt
\parskip 0pt plus 5pt

~\vglue2cm
\begin{center}
\Large\bf
Searching for Quark Matter at the CERN SPS
\end{center}

\bigskip\bigskip

\begin{center}
C. Louren\c{c}o

\bigskip

CERN-EP, CH-1211 Geneva 23, Switzerland

\vspace{2.5cm}
\textbf{Abstract}
\end{center}

\begingroup
\leftskip=0.4cm
\rightskip=0.4cm
\parindent=0.pt

This article gives a brief overview of some recent advances in our
understanding of the physics of dense strongly interacting matter,
from measurements done at the CERN SPS.  The presently available
results are very interesting, and are likely to reflect the production
of a new state of matter in central Pb-Pb collisions, at the highest
SPS energies.  However, important questions require further work.
Particular emphasis is given to developments made since the Quark
Matter 1999 conference, and to issues that justify the continuation of
the SPS heavy ion physics program beyond year 2000.

\vspace{2.5cm}

\begin{center}
\emph{Invited talk presented at the Quark Matter 2001 conference\\
Stony Brook, USA, January 14-20, 2001\\ 
To be published in the Proceedings}
\end{center}

\endgroup

\clearemptydoublepage
\pagenumbering{arabic}
\setcounter{page}{1}

\title{Searching for Quark Matter at the CERN SPS}

\author{Carlos Louren\c{c}o\\
\vspace{2mm}
CERN-EP, Geneva, Switzerland}

\maketitle

\begin{abstract}

This article gives a brief overview of some recent advances in our
understanding of the physics of dense strongly interacting matter,
from measurements done at the CERN SPS.  The presently available
results are very interesting, and are likely to reflect the production
of a new state of matter in central Pb-Pb collisions, at the highest
SPS energies.  However, important questions require further work.
Particular emphasis is given to developments made since the Quark
Matter 1999 conference, and to issues that justify the continuation of
the SPS heavy ion physics program beyond year 2000.

\end{abstract}

\section{Introduction}

The study of high energy heavy ion collisions is presently a very
active field in experimental particle physics, with the RHIC collider
in operation at BNL since summer 2000 and with the ALICE experiment
being prepared to study this kind of physics at the LHC energies.  The
first goal of this experimental program, which started in 1986, with
the AGS and SPS fixed-target programs, is the discovery of the phase
transition from confined hadronic matter to deconfined partonic
matter.  The idea that such a phase transition should exist, between
hadronic and quark matter, is around since the first models of the
quark structure of hadrons.  It is presently studied in detail in the
framework of Lattice QCD calculations, which predict its occurrence
when the temperature of the system exceeds a critical threshold at
around 170~MeV, corresponding to a critical energy density of around
700~MeV/fm$^3$~\cite{qm01-fk}.  The proof of existence of the quark
matter phase and the study of its properties are key issues in QCD,
for the understanding of confinement and chiral-symmetry.

When this new state of matter was postulated, some signatures of its
formation in high energy nuclear collisions were proposed, on the
basis of theoretical arguments, among which we can highlight the
enhancement of strange particle production, the suppression of
charmonia states (\jpsi, $\chi_c$ and \psip), due to the screening of
the \ccbar\ binding potential in the QGP colour soup, and the
production of thermal dileptons, electromagnetic radiation emitted by
the `free' quarks.  The results obtained by the SPS experiments, after
15 years of collecting data with proton and ion beams, provide
``compelling evidence for the existence of a new state of matter, in
which quarks roam freely'', produced in central Pb-Pb collisions at the
highest SPS energies.  Among the most exciting observations are the
enhanced production of multistrange hyperons, the centrality
dependence of the \jpsi\ suppression pattern and the enhancement of
intermediate mass dimuon production, a possible indication of thermal
dimuons.  Besides, the enhanced production of low mass dileptons may
be an indication of approach to chiral symmetry restoration.

In view of these exciting results and in order to clarify important
questions remaining open, a new experiment, NA60, has been approved at
CERN, extending the SPS runs with heavy ion beams and bridging the gap
between the original SPS program and the future high-energy wonderland
of ALICE.  At the other end of the energy scale, the NA49 experiment
will have a ten day extension with Pb ions at 20 and 30~GeV per
nucleon, to complete the energy scan of global strangeness production.
So far it has collected data at 158, 40 and 80~GeV per nucleon.  This
extension of the SPS heavy ion running time will bring to a proper
conclusion the program started in 1986, whilst providing valuable
information, complementary to the studies underway at RHIC.

This paper gives a somewhat personal overview of the present status of
the SPS results, briefly mentions the progress made since the last
Quark Matter conference, presents some pertinent questions that remain
open, and dreams about how the future SPS program could address those
issues.

\section{Present status of the SPS results}

The very large amount of experimental results obtained by the CERN SPS
experiments since 1986 is so vast and diversified that a proper review
would require a much more extensive article, jointly prepared by
several of the active players in the field.  The previous conference
in the Quark Matter series, the last one before the RHIC experiments
started collecting data, ended with two summary talks that reviewed in
detail the status of the field in terms of hadronic~\cite{qm99-rs} and
dilepton~\cite{qm99-lk} signals.  A few attempts have also been made
to see in a coherent way some of the most significant results, in
particular those obtained with the lead beam at the SPS~\cite{rs-uh}.
A tentative summary has been proposed~\cite{cernpr}, basically saying
that ``the combined results provide compelling evidence for the
existence of a new state of matter, featuring many of the
characteristics of the primordial soup in which quarks and gluons
existed before they clumped together as the universe cooled down''.
The present paper is too short, and is not the right place, for
critically commenting the scientific meaning (and opportunity) of
these words.

It is certainly appropriate to say that all the CERN SPS experiments
have been successful in delivering significant information, many of
them having seen ``what they were looking for''.  However, these 15
years have also confirmed that heavy ion collisions lead to very
complicated (and fastly evolving) systems, and that it is difficult to
extract clear messages from the observations.  Central collisions
between two Pb nuclei, at the highest SPS energies, lead to the
production of hundreds of final-state particles widely emitted without
discernible structures.  These complex, and apparently chaotic, final
states can be studied applying statistical concepts, attempting
descriptions based on (non-perturbative) QCD thermodynamics, and
summarized by macroscopic variables like temperature, pressure, etc.
But such studies present formidable challenges, requiring complex (and
quite expensive) experimental techniques, demanding huge amounts of
computing time, and leading, after a major effort of many people, to a
few points on a figure.  The usefulness of such measurements must be
evaluated by their ability to rule out existing alternative models or,
at least, by their power in constraining assumptions and free
parameters.

One of the most important lessons from the SPS program has been, in
fact, that it is extremely difficult to rule out ``continuously
improving models'' that are always able to reproduce, \emph{a
posteriori}, any new measurements.  In these conditions, the fact that
``we have seen what we were looking for'', does not mean that we have
made any progress in our understanding.  Twenty years after the first
Quark Matter conferences, and after a considerable evolution in the
theoretical and experimental sectors of this field, we still wonder,
frequently, if we really know ``what we are looking for''.  In spite
of the strong indications that very interesting phenomena occur in the
early stages of a Pb-Pb collision, at the highest SPS energies, we
still do not know the answer to the critical question that motivates
this field: can we convince ourselves and the community at large that
we have formed quark matter in the laboratory?

In our field, as in most scientific enterprises, each step forward
brings new questions, at the same time as it solves older ones.  The
value of a scientific discovery is, also, in the relevance of the new
questions it raises.  But have we really solved any question so far?
How many models, that clearly fail to properly reproduce already
established results, using parameters within reasonable ranges, have
been definitely ruled out?  Once more, in this conference, we have
seen old models, with new free parameters, in ridiculous exercises
attempting to describe experimental measurements, bluntly ignoring
their (small) error bars, and just succeeding in increasing the level
of entropy in the field.  We should be happy when our models are
successfully falsified.  As Bo Andersson said in his Quark Matter 1985
paper~\cite{qm85-ba}, quoting Bacon, ``nature never tells you when you
are right but only when you are wrong.  Therefore, you have only
learned something when you disagree with the data.  For some reason,
the scientific society does not seem to honour the successfully wrong
theorists''.  While looking for `critical phenomena' in nature, we
must follow a critical behaviour in our own work, both in the
measurement and in the interpretation steps.  ``Whenever we propose a
solution to a problem, we ought to try as hard as we can to overthrow
our solution, rather than defend it.  Few of us, unfortunately,
practice this precept; but other people, fortunately, will supply the
criticism for us''~\cite{popper}.

The final clarification of the present SPS results requires a careful
and systematic approach, to establish beyond reasonable doubt that the
QCD phase transition from hadronic to quark matter happens in central
Pb-Pb collisions at the highest SPS energies.  The available
measurements should be critically scrutinized, looking for possible
biases of any kind.  In cases where information is obviously missing,
new measurements should be urgently done.  Models that pretend to
explain the available results must provide specific predictions for
future measurements, with appropriate and carefully explained
uncertainty bands.  If the new observations validate those
predictions, we will have, then, made substantial progress.
Otherwise, without a devoted and objective effort, seriously joining
theory and experiment, the epitaph of the SPS heavy ion program will
be ``just when we were about to find the answer, we forgot the
question''.

\section{Strangeness production}

One of the earliest predictions in the field is that particles
containing strange quarks should be produced more often if the system
produced in the heavy ion collisions passes through a quark-gluon
plasma phase.  A global increase of around a factor 2 in strangeness
production (dominated by kaon production), has indeed been observed,
in particular by the NA49 large-acceptance experiment.  The most
spectacular observations have been done, however, in the multi-strange
hyperon sector.  The very large enhancement factors in particle yields
per participating nucleon, reaching a factor around 17 for the
$\Omega$, a triple strange hyperon, and the fact that these factors
are significantly higher for the states with more strange quarks,
i.e.\ $E_{\Omega}>E_{\Xi}>E_{\Lambda}$, where $E_i$ is the enhancement
of the particle $i$ with respect to p-A interactions, are naturally
explained if the particle yields are determined from statistical
hadronization of a strangeness-enhanced plasma phase.  On the
contrary, such enhancement levels cannot be reproduced in conventional
(final state hadronic rescattering) scenarios, given the short
lifetime of the expanding hadronic system (see, however,
Ref.~\cite{qm01-kr}).  When these results were presented in the Quark
Matter 1999 conference~\cite{qm99-fa}, a question was left in the air:
is there a threshold behaviour in the enhancement pattern, between the
p-Be and p-Pb points and the Pb-Pb values?  The flat pattern observed
in the Pb-Pb data indicated very little dependence on the centrality
of these collisions, for $N_\mathrm{part}>100$.  Where was the
transition?

The NA57 experiment has continued these studies, making a special
effort to collect peripheral Pb-Pb collisions.  The first results,
presented at this conference~\cite{qm01-na57}, show that the
enhancement of $\bar{\Xi}^+$ production increases by a factor 2.6,
from $N_\mathrm{part}=62$ to 121.  The confirmation of a threshold
behaviour in the strangeness enhancement pattern may come from the
data analysis of the other strange hyperons.  Unfortunately, having
only five bins in centrality, and four of them showing a flat
behaviour, the Pb-Pb pattern measured by the NA57 experiment will fall
short of showing a clear transition, with a characteristic threshold,
if it exists.  A better understanding of the apparent onset of the
enhancement would require studying ``intermediate mass'' nuclear
collisions, like In-In, for instance, in small centrality steps.
Unfortunately, the effort required by the preparation of the ALICE
experiment seems to prevent the realisation of such future studies.

Another major addition to the strangeness chapter is being provided by
the NA49 experiment, when running with proton beams.  The very large
acceptance of this experiment makes it particularly appropriate in the
study of asymmetric collision systems, as p-A interactions, where the
reflection of the probed phase space window around midrapidity cannot
be performed.  First results of ``NA49-hadrons'' have been shown at
this conference~\cite{qm01-na49}, raising some questions on the ``p-A
reference baseline'' used by WA97/NA57 in the extraction of the
enhancement factors.  Fortunately, we will see further data on this
issue in the near future, since ``NA49-hadrons'' has been approved for
further running in the next few years, and NA57 will collect more data
on proton induced collisions in 2001.

Still in the strangeness sector, long standing questions concerning
$\phi$ production remain unclear.  NA50 sees, in the dimuon decay
channel, a strong increase in the yield of $\phi$ mesons produced in
heavy ion collisions and a transverse mass spectrum with a rather low
``inverse slope''~\cite{na50-phi}, contrary to the observations of the
NA49 experiment~\cite{na49-phi}, in the K$^+$K$^-$ decay channel.
Future measurements of low-\pt\ $\phi$ production in the dilepton
channel, by NA60, should help clarifying the source of discrepancy.

\section{Low mass dilepton production}

The CERES experiment has observed that the yield of low mass $e^+e^-$
pairs measured in p-Be and p-Au collisions is properly described by
the expected ``cocktail'' of hadronic decays, while in \mbox{Pb-Au}
collisions, on the contrary, the measured yield, in the mass region
0.2--0.7~GeV, exceeds by a factor 2.5 the expected
signal~\cite{qm99-na45}.  These observations are consistent with the
expectation that the properties of vector mesons should change when
produced in dense matter.  In particular, near the phase transition to
the quark-gluon phase, chiral symmetry should be partially restored,
making the vector mesons indistinguishable from their chiral partners,
thereby inducing changes in their masses and decay widths.  The short
lifetime of the $\rho$ meson, shorter than the expected lifetime of
the dense system produced in the SPS heavy ion collisions, makes it a
sensitive probe of medium effects and, in particular, of chiral
symmetry restoration.

Already E.~Shuryak~\cite{qm99-es} and B.~M\"uller~\cite{qm99-bm}, in
their Quark Matter 1999 papers, emphasized the importance of a
considerable improvement in the CERES measurements of low mass
dilepton production, in terms of signal to background ratio, mass
resolution, and statistics.  A TPC was added to the CERES setup, to
improve the momentum resolution of the dielectron measurement.
Unfortunately, problems in the data taking during year 1999 have
prevented the CERES experiment from collecting a reasonable sample of
dilepton events~\cite{qm01-na45}.  Those problems were solved in time
for the run of year 2000, but a first look into this new data sample
indicates that the anticipated performance has not been fully reached,
at least in what concerns mass resolution and statistics.  An accurate
measurement of the $\omega$ resonance, necessary as a reference in the
studies of the in-medium modifications apparently affecting the
$\rho$, may remain in the wish list until the first runs of the NA60
experiment.

\section{Intermediate mass dilepton production}

The NA38 and NA50 experiments have studied the production of dileptons
in the mass window between the $\phi$ and the \jpsi\ peaks, as a
superposition of Drell-Yan dimuons and simultaneous semileptonic
decays of $D$ and $\bar{D}$ mesons, after subtraction of the
combinatorial background from pion and kaon decays~\cite{qm01-lc}.
The dimuon mass spectra measured in p-A collisions are very well
reproduced taking the high mass region to normalize the Drell-Yan
component and an open charm cross-section in good agreement with
direct measurements made by other experiments.  On the contrary, the
superposition of Drell-Yan and open charm contributions, with the
nucleon-nucleon absolute cross sections scaled with the product of the
mass numbers of the projectile and target nuclei (as expected for hard
processes), fails to properly describe the dimuon yield measured in
ion collisions.

The data can be properly reproduced by simply increasing the open
charm yield, with a scaling factor that grows linearly with the number
of nucleons participating in the collision, reaching a factor 3 in the
most central Pb-Pb collisions.  The observed excess can also be due to
the production of thermal dimuons, a signal that was the original
motivation for the NA38 experiment.  In particular, the intermediate
mass dimuons produced in the most central Pb-Pb collisions are well
reproduced by adding thermal radiation, calculated according to the
model of Ref.~\cite{rapp}, to the Drell-Yan and charm contributions
normally extrapolated from nucleon-nucleon collisions.  This model
explicitly includes a QGP phase transition with a critical temperature
of 175~MeV.  The best description of the data is obtained using
$\sim$\,250~MeV as the initial temperature of the QGP medium radiating
the virtual photons.  The presently available data cannot distinguish
between an absolute enhancement of charm production and the emission
of thermal dilepton radiation.  The clarification of the nature of the
physics process behind the observed excess is also in the wish lists
presented by E.~Shuryak and B.~M\"uller in their Quark Matter 1999
papers, and is the strongest physics motivation of the NA60
experiment.

\section{Charmonia production and suppression}

The formation of a deconfined medium should induce a considerable
suppression of the charmonia production rate, due to the colour
`Debye' screening of the $c\bar{c}$ potential or to the breaking of
the $c\bar{c}$ binding by scattering with energetic (deconfined)
gluons.  However, even the relatively simple measurement of \jpsi\
production faces a big challenge when it comes to furnish a convincing
logical case that proves, to the satisfaction of the experts in the
field, that a deconfined state of matter has been formed.  It is not
enough to show that a certain observable changes from p-Pb to Pb-Pb
collisions, for instance, or to argue that its value in the most
central nucleus-nucleus collisions is different from what is
calculated in a `conventional physics' model.  The best path to
clearly establish a solid result and shed light in this complicated
field is to build a robust set of measurements, that establishes a
precise reference baseline, relative to which the specific behaviour
of heavy ion collisions can be extracted.  Such a baseline shows what
is the `normal' behaviour of the signal we are studying, with respect
to which we look for changes due to QGP formation.  Furthermore, we
are in a much better position if nature provides us with a reference
process, insensitive to the formation of a deconfined phase, specially
if we can measure it with the same detector.

In the case of the \jpsi\ suppression topic, the baseline is built
from the measurements done by NA38 and NA50 with pp, p-A and light ion
collisions~\cite{qm01-es}.  Very peripheral Pb-Pb collisions have been
successfully collected in year 2000 and we will soon know how well
they follow the ``normal nuclear absorption'' baseline.  The best
reference physics process, at SPS energies, is the rate of high mass
Drell-Yan dimuons, seen to scale with the number of binary nucleon
collisions, up to the Pb-Pb case.  It is well known that the \jpsi\
production yield, in Pb-Pb collisions, shows a very peculiar behaviour
as a function of the centrality of the collisions, estimated either
using the transverse or the forward energies~\cite{qm01-pb}.  The
observed two-step suppression pattern is in clear disagreement with
the predictions of `conventional' models, which attribute the
disappearance of the \jpsi\ mesons to interactions with `comoving'
hadrons, while it is naturally explained if the charmonia states are
dissolved in a deconfined medium, due to the different melting
temperatures of the $\chi_c$ and \jpsi\ states (about 30--40\,\% of
the observed \jpsi\ mesons result from the decay of $\chi_c$ states).

If we accept that this pattern indicates the production of a state of
matter where colour is no longer confined, we must move on to the
detailed understanding of how deconfinement sets in, and what physics
variable governs the threshold behaviour of the ($\chi_c$)
suppression: (local) energy density, density of wounded nucleons, of
percolation clusters, of produced gluons, etc.  This requires
collecting data with a smaller nuclear collision system like
\mbox{In-In}.  Indeed, it is possible to predict at which impact
parameter, $b$, of \mbox{In-In} collisions is reached the same
threshold in (local) energy density, or any other variable, as reached
in \mbox{Pb-Pb} collisions of $b\approx 8$~fm, where the $\chi_c$
state starts melting.  A verification of such specific predictions
would be the final element of proof that the deconfined quark-gluon
phase sets in, and would provide fundamental information on the
mechanisms behind the observed phenomena.  In this context, it is also
important to improve our knowledge of the nuclear dependence of
$\chi_c$ production, in p-A collisions, at SPS energies.

The \jpsi\ data do not provide a direct measurement of the critical
temperature.  Finite temperature lattice QCD tells us that the
strongly bound \jpsi\ \ccbar\ state should be screened when the medium
reaches temperatures 30--40\,\% higher than $T_c$, while the larger
and more loosely bound \psip\ state should melt near $T_c$.  The
\psip\ is already significantly suppressed when going from p-U to
peripheral S-U collisions but the presently existing results are not
clear in what concerns the pattern of the \psip\ suppression.  Is the
``anomalous'' \psip\ suppression due to QGP melting or to hadronic
absorption?  If we see that this suppression happens more or less in
an abrupt way, within a single collision system rather than comparing
p-U to S-U data, we would know that Debye screening is the mechanism
responsible for the \psip\ disappearance and we would have a clear
measurement of $T_c$.  This requires a new measurement, with improved
mass resolution to have a cleaner separation between the \psip\ and
\jpsi\ peaks, and which scans an energy density region including the
p-U and the S-U points.

Improved measurements of \jpsi\ and \psip\ production, with
intermediate mass nuclei, were also included in the wish lists of
E.~Shuryak and B.~M\"uller, and are an important part of the physics
program of the NA60 experiment, which will also measure $\chi_c$
production in p-A collisions.

\section{Open charm production}

Knowing that the bound $c\bar{c}$ states are suppressed, it is natural
to ask what happens to the \emph{unbound} charm.  Charm quarks are so
heavy that they can only be produced at the earlier stages of the
nuclear collision, before the eventual formation of the QGP state.
Charm is the heaviest flavour that can be studied in heavy ion
collisions at the SPS energies.  The production of charm quarks leads
mainly to correlated pairs of $D$ and $\bar{D}$ mesons.  Only a few
percent of the charmed quark pairs end up in the bound charmonia
states presently studied by the NA50 experiment, and which exhibit a
rather interesting ``anomalous'' behaviour.  What happens to the vast
majority of $c$ quarks?  Are they affected by energy loss while
crossing the dense (partonic or hadronic) medium?  Is charm production
enhanced similarly to what has been seen in the strangeness sector?
Finally, $D$ meson production provides the natural reference with
respect to which we should study the observed \jpsi\ suppression,
since both production mechanisms depend on the same gluon distribution
functions.  If charm production is enhanced in nuclear collisions, it
makes the \jpsi\ suppression even more anomalous.  A direct
observation of $D$ meson production is clearly the most important new
measurement that remains to be done at the SPS, and constitutes a
basic reason for the construction and running of NA60.

\section{Future prospects}

The results and open questions presented in the previous sections
emphasize the importance of having a new experiment at the SPS, that
can significantly improve several existing observations and make a few
new measurements, including a measurement of open charm production in
heavy ion collisions.  A dedicated experiment is needed, that can cope
with the very high particle multiplicities reached in the most central
nuclear collisions (400 charged particles per unit rapidity at
midrapidity) and with the rather small $D$ production cross section.

The NA50 experiment has been using CERN's highest intensity heavy ion
beam (more than $10^7$~ions per second) and has a very selective
dimuon trigger, quite appropriate to look for rare processes.  The
recently approved NA60 experiment~\cite{na60}, complements the muon
spectrometer and zero degree calorimeter already used in NA50 with two
state-of-the-art silicon detectors, placed in the target region: a
radiation hard beam tracker, consisting of four silicon microstrip
detectors placed on the beam and operated at a temperature of 130~K,
and a 10-plane silicon pixel tracking telescope, made with radiation
tolerant readout pixel chips, placed in a 2.5~T dipole magnetic field.

The NA60 experiment has been approved to run from 2001 to 2003, using
proton, Pb and In beams.  The following questions summarize the
physics motivation of NA60.
\begin{itemize}
\parskip=0.pt \parsep=0.pt \itemsep=0.pt
\item What is the origin of the dimuon excess seen in the intermediate
mass region?  Thermal dimuon production?
\item Is the open charm yield enhanced in nucleus-nucleus collisions?
How does it compare to the suppression pattern of bound charm states?
\item What is the variable (local energy density, cluster density,
etc.) that rules the onset of charmonia suppression?
\item What is the physical origin of the \psip\ suppression?  If it is
due to Debye screening, what is its melting temperature?
\item Which fraction of the \jpsi\ yield comes from $\chi_c$ decays?
Does it change from p-Be to p-Pb collisions?
\item Are there medium induced modifications in the $\rho$ meson?  Is
there a threshold behaviour in the low mass dilepton enhancement?
What happens with the $\omega$ meson?
\item Is the observed $\phi$ enhancement a specific feature of heavy
ion collisions?  Is the $\phi$ sensitive to flow?
\end{itemize}

The beam tracker gives the transverse coordinates of the interaction
point, on the targets, with enough accuracy to measure the impact
parameter of the muon tracks, i.e.\ the minimum distance between the
track and the collision vertex, in the transverse plane.  Thanks to
this information, NA60 will be able to separately study the production
of prompt dimuons and the production of muons originating from the
decay of charmed mesons, in p-A and heavy ion collisions.  The prompt
dimuon analysis will use events where both muons come from (very close
to) the interaction vertex.  The open charm event sample is composed
of those events where both muon tracks have a certain minimum offset
with respect to the interaction point and a minimal distance between
themselves at $z_\mathrm{vertex}$.  It should not be difficult to see
which of these two event samples is enhanced by a factor of 2 or 3 in
nuclear collisions of $N_\mathrm{part}\sim 300$.

\begin{figure}[ht]
\centering
\begin{tabular}{cc}
\resizebox{0.48\textwidth}{!}{%
\includegraphics*{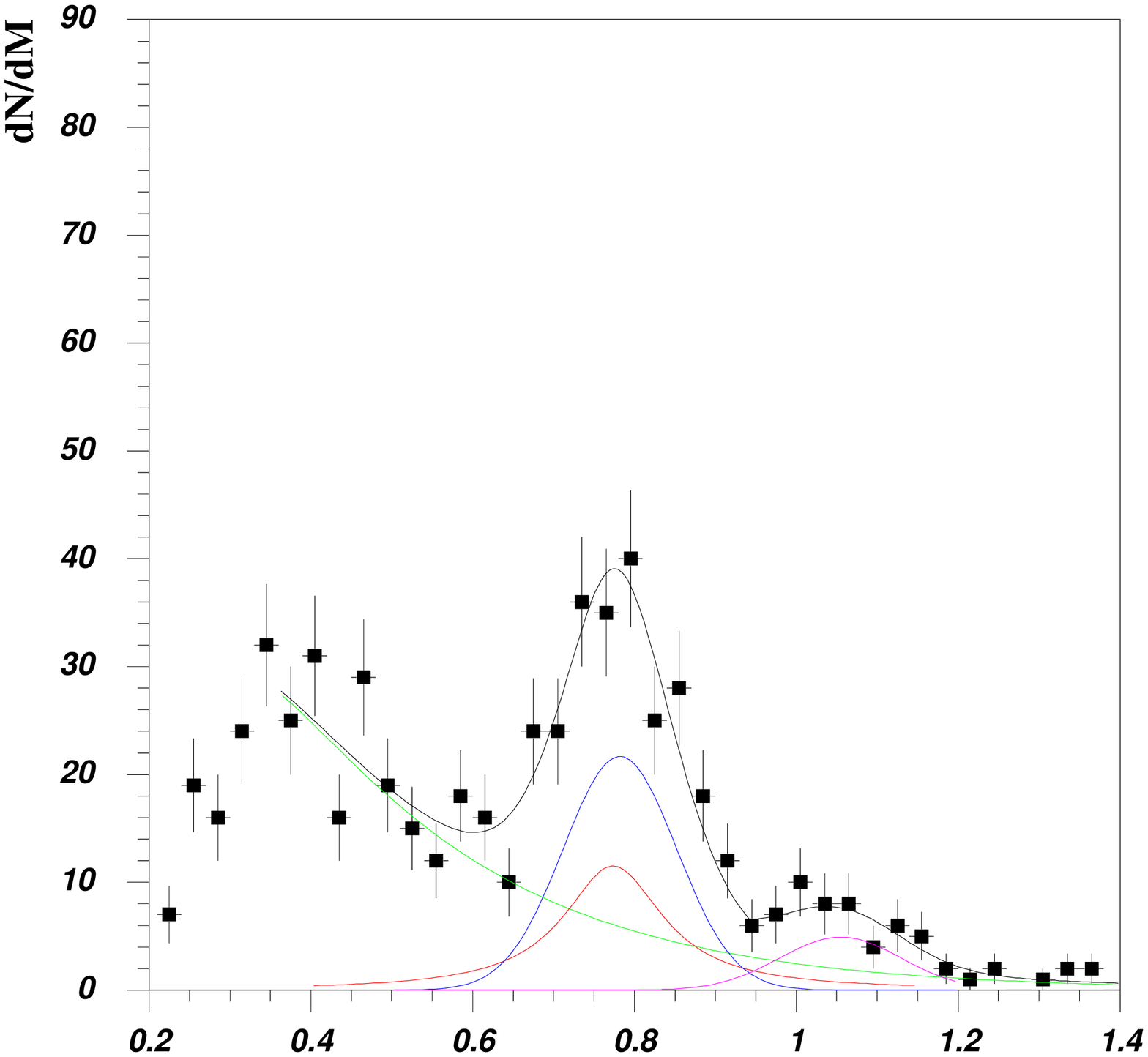}}
&
\resizebox{0.48\textwidth}{!}{%
\includegraphics*{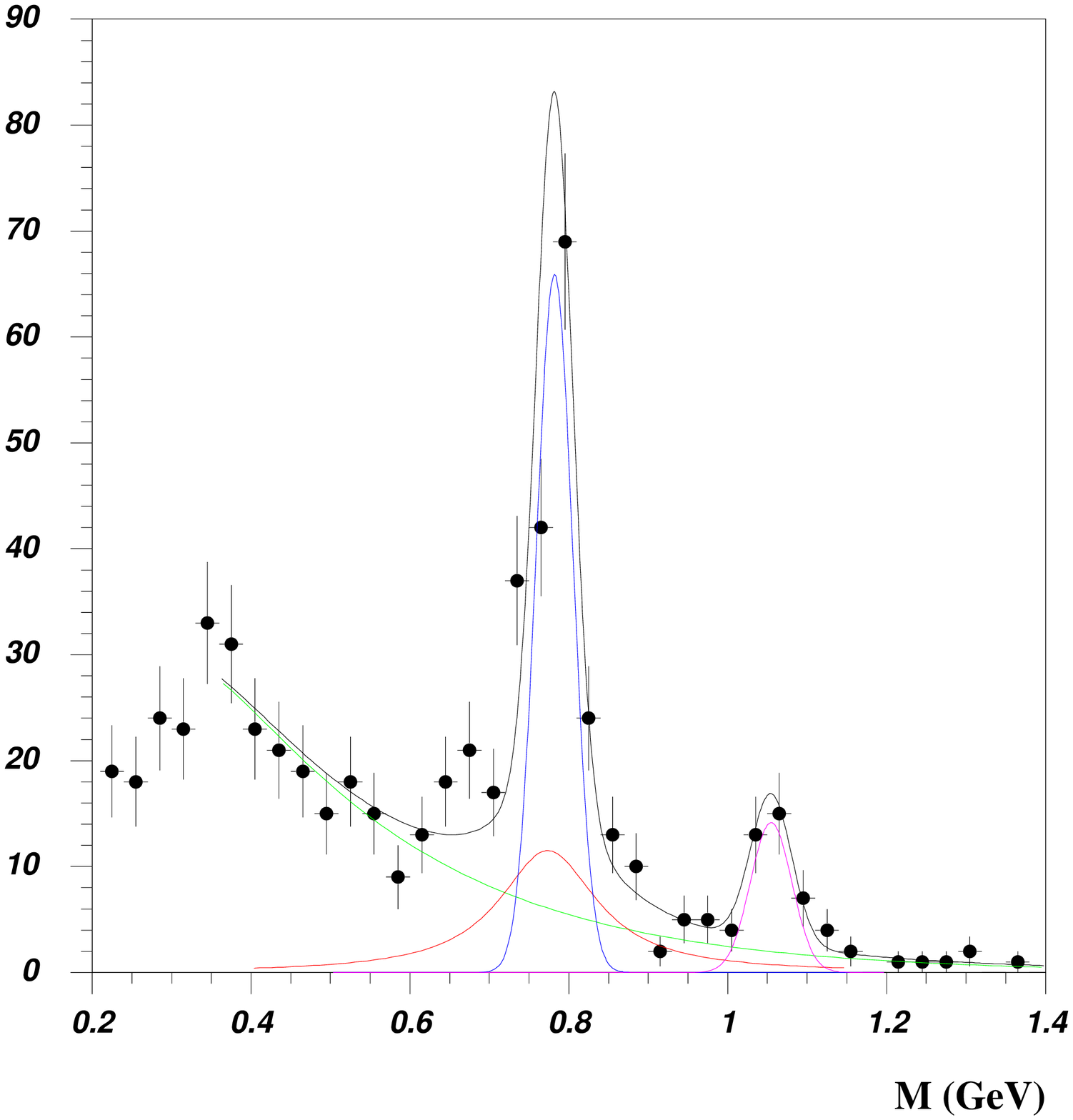}}
\end{tabular}
\vglue-0.4cm
\caption{Dimuon mass distributions measured in 1998, in p-Be
collisions, before (left) and after (right) using the information of
the test pixel telescope.  The curves represent the low mass vector
meson resonances ($\rho$, $\omega$ and $\phi$) on the top of a
continuum.  They are normalized to the same number of events in both
figures.  The collected statistics (600 events) correspond to a few
minutes of NA60 running.}
\label{pbe}
\end{figure}

The high granularity tracking telescope, placed in a powerful dipole
field, gives access to the muon tracks at the vertex level and vastly
improves the mass resolution of the dimuon measurement.  This has been
demonstrated in a very fast feasibility test done in 1998, using a
small telescope (four half-planes) made of the previous generation of
readout pixel chips.  The results, shown in Fig.~\ref{pbe}, confirm
that the mass resolution improves from 70 to 20~MeV at the $\omega$
mass, as expected from the physics performance simulations.

The studies of prompt dimuon and open charm production in \mbox{p-A}
collisions are important reference measurements, to understand the
results obtained with nuclear collisions.  In particular, the ratio
between the open charm and the Drell-Yan production cross sections
will be determined with high accuracy in several \mbox{p-A} collision
systems, revealing if these two hard processes have the same
A-dependence or not.  More challenging will be the measurement of the
dependence of $\chi_c$ production on the mass number of the target, in
p-A collisions, by seeing how the ratio between $\chi_c$ and \jpsi\
yields changes from \mbox{p-Be} to \mbox{p-Pb} collisions.  To
minimize the systematical uncertainties due to the beam flux
normalization, the measurement will be done using the Be and Pb
targets simultaneously in the beam.  The $\chi_c \to \psi~ \gamma \to
\psi~ e^+e^-$ decays will be used for this study, with the photons
converting in a Pb disk placed downstream of the targets, and the
electron-positron pairs reconstructed in the silicon pixel telescope.

\section{Summary and conclusions}

Starting from the questions and wishes expressed at the Quark Matter
1999 conference, concerning measurements that should be done at the
CERN SPS before closing this facility, I have briefly mentioned some
of the most recent developments and emphasized the issues that have
imposed a continuation of the SPS heavy ion physics program.

After 15 years of ``learning curve'', we can say that we have been
unable to falsify the hypothesis of quark gluon plasma formation at
the CERN SPS.  In fact, as predicted, strangeness is enhanced, the
\jpsi\ is suppressed, the dimuon continuum looks as if thermal
dileptons are produced, and there are modifications in the low mass
dilepton spectra, among other observations.  These results provide
extremely relevant information about the (predicted) formation of a
deconfined state of matter in high energy heavy ion collisions.
However, considerable homework remains to be done in view of
converting ``compelling evidence'' into ``conclusive evidence'' that
the quark matter phase has indeed been formed at CERN.  This is
exactly the reason why the heavy ion community must make a significant
effort to further clarify the present results and reach a deeper
understanding of the critical behaviour of QCD at SPS energies.

The re-birth of the heavy ion physics program at the CERN SPS, with
the extension of NA49 and the approval of the new NA60 experiment,
represents an evolution from a broad physics program to a dedicated
study of specific signals that already provided very interesting
results.  It is not clear that the NA60 experiment will run
successfully, given the delays in the availability of pixel readout
chips and the dramatic lack of resources (people and budget).  If it
does, the new measurements should give a significant contribution to
the understanding of the presently existing results, and considerably
help in building a convincing logical case that establishes beyond
reasonable doubt the formation (or not) of a deconfined state of
matter in heavy ion collisions at the SPS.


\begin{thebibliography}{99}
\parskip=0.pt \parsep=0.pt \itemsep=0.pt

\bibitem{qm01-fk} F. Karsch, these proceedings.

\bibitem{qm99-rs} R. Stock, Proc. of QM'99,
\Journal{\NPA}{661}{282c}{1999}.

\bibitem{qm99-lk} L. Kluberg, Proc. of QM'99,
\Journal{\NPA}{661}{300c}{1999}.

\bibitem{rs-uh} R. Stock, \Journal{\PLB}{456}{277}{1999}\\ 
U. Heinz, Proc. of QM'99, \Journal{\NPA}{661}{140c}{1999}.

\bibitem{cernpr} M. Jacob and U. Heinz, CERN press release, Feb. 2000.

\bibitem{qm85-ba} B. Andersson, Proc. of QM'85,
\Journal{\NPA}{447}{165c}{1985}.

\bibitem{popper} K. Popper, ``The Logic of Scientific Discovery'',
1959.

\bibitem{qm01-kr} K. Redlich, these proceedings.

\bibitem{qm99-fa} F. Antinori \emph{et al.} (WA97 Coll.), Proc. of
QM'99, \Journal{\NPA}{661}{130c}{1999}.

\bibitem{qm01-na57} N. Carrer \emph{et al.} (NA57 Coll.), these
proceedings.

\bibitem{qm01-na49} T. Susa \emph{et al.} (NA49 Coll.), these
proceedings.

\bibitem{na50-phi} N. Willis \emph{et al.} (NA50 Coll.), Proc. of
QM'99, \Journal{\NPA}{661}{534c}{1999}.

\bibitem{na49-phi} V. Friese \emph{et al.} (NA49 Coll.), these
proceedings.

\bibitem{qm99-na45} B. Lenkeit \emph{et al.} (CERES Coll.), Proc. of
QM'99, \Journal{\NPA}{661}{23c}{1999}.

\bibitem{qm99-es} E. Shuryak, Proc. of QM'99,
\Journal{\NPA}{661}{119c}{1999}.

\bibitem{qm99-bm} B. M\"uller, Proc. of QM'99,
\Journal{\NPA}{661}{272c}{1999}.

\bibitem{qm01-na45} H. Appelshauser \emph{et al.} (CERES Coll.), these
proceedings.

\bibitem{qm01-lc} L. Capelli \emph{et al.} (NA50 Coll.), these
proceedings.

\bibitem{rapp} R. Rapp and E. Shuryak, \Journal{\PLB}{473}{13}{2000}.

\bibitem{qm01-es} E. Scomparin \emph{et al.} (NA50 Coll.), these
proceedings.

\bibitem{qm01-pb} P. Bordalo \emph{et al.} (NA50 Coll.), these
proceedings.

\bibitem{na60} A. Baldit \emph{et al.} (NA60 Coll.), Proposal
CERN/SPSC 2000-010, March 2000.

\end{thebibliography}
\end{document}